\documentclass[journal]{IEEEtran}
\usepackage{psfrag}
\usepackage{amsmath}
\usepackage{subfigure}
\usepackage{url}
\usepackage{stfloats}
\usepackage{cite}
\usepackage{amsmath}
\usepackage{mathtools}
\usepackage{upgreek}
\usepackage{amssymb}
\usepackage{amsfonts}
\usepackage{graphicx}
\usepackage{fancybox}
\usepackage{color}
\usepackage{algorithm}
\usepackage{algorithmic}
\usepackage{multirow}
\usepackage{booktabs}
\usepackage{threeparttable}
\usepackage{latexsym}
\usepackage{bm,array}
\usepackage{graphicx}
\usepackage{float}
\usepackage{hyperref}
\usepackage{acronym}
\usepackage{textcomp}

\IEEEoverridecommandlockouts
\normalsize

\begin{document}
\title{\textcolor{black}{Deep Q-Network Based Dynamic Movement Strategy in a UAV-Assisted Network} \vspace{0.2cm}}
\author{Xukai Zhong, Yiming Huo,~\IEEEmembership{Member,~IEEE}, Xiaodai Dong,~\IEEEmembership{Senior Member,~IEEE}, and Zhonghua Liang,~\IEEEmembership{Senior Member,~IEEE}
\thanks{X. Zhong, Y. Huo and X. Dong are with the Department of Electrical and Computer Engineering, University of Victoria, Victoria, BC V8P 5C2, Canada (e-mail: xukaiz@uvic.ca, ymhuo@uvic.ca, xdong@ece.uvic.ca). This work was supported by Wighton Engineering Product Development Fund. (corresponding author:\emph{~Xiaodai Dong})} 
 \thanks{Z. Liang is with School of Information Engineering, Chang’an University, Xi’an, Shanxi Province, China (e-mail: lzhxjd@hotmail.com). Z.  Liang’s work was supported in part by  the Natural Science Basic Research  Project in Shaanxi Province of China under Grant 2020JM-242, and in part by the Fundamental Research Funds for the Central Universities, CHD under Grant 300102249303.} 

\vspace{-0.4cm}} 

\maketitle

\begin{abstract}
Unmanned aerial vehicle (UAV)-assisted communications is a promising solution to improve the performance of future wireless networks, where UAVs are deployed as base stations for enhancing the quality of service (QoS) provided to ground users when traditional terrestrial base stations are unavailable or not sufficient. An effective framework is proposed in this paper to manage the dynamic movement of multiple unmanned aerial vehicles (UAVs) in response to ground user mobility, with the objective to maximize the sum data rate of the ground users. First, we discuss the relationship between the air-to-ground (A2G) path loss (PL) and the location of UAVs. Then a deep Q-network (DQN) based method is proposed to adjust the locations of UAVs to maximize the sum data rate of the user equipment (UE). Finally, simulation results show that the proposed method is capable of adjusting UAV locations in a real-time condition to improve the QoS of the entire network.        
\end{abstract}


%
\IEEEpeerreviewmaketitle

\vspace{-0.3cm}
\section{Introduction}

\IEEEPARstart{T}{he} unprecedented demand for high-quality wireless communications has fueled the evolution of wireless technologies and communications networks. The unmanned aerial vehicle (UAV)-assisted network where UAVs are deployed and function as aerial base stations to assist the terrestrial base stations is an effective complementary solution to emergency wireless service recovery after natural disasters or infrastructure damage~\cite{Zeng2016}. Also, in Internet of Things (IoT) networks, UAVs can be used as aerial base stations to collect data from ground devices, in which building a complete cellular infrastructure is not \textcolor{black}{affordable}~\cite{Mozaffari2017a}. The authors in~\cite{Huo2019} \textcolor{black}{proposed} a multi-layer UAV network model for UAV-enabled 5G and beyond applications. Despite advantages such as flexibility, mobility, cost and time efficiency in UAV-assisted networks, one key design challenge is to determine the move strategy for UAVs. Since in realistic situations, the environment where UAVs are deployed is highly dynamic, it is critical for UAVs to adjust its locations regularly to cope with varying conditions. \textcolor{black}{Furthermore, utilizing machine learning techniques for the UAV communication recently has seen unprecedented growing popularity~\cite{Bithas}.}

\subsection{Related Work}
Regarding the existing research related to the UAVs deployment, the authors in~\cite{Mozaffari2016} modeled the static UAV deployment problem based on circle packing theory and studied the relationship between the number of deployed UAVs and the coverage time. Moreover,~\cite{Al-Hourani2014} proposed an efficient air-to-ground (A2G) channel model with probabilistic path loss (PL) and discussed a method to derive the optimal altitudes of UAVs based on the A2G channel model. The work in~\cite{Shi2017} discussed an optimization problem in the UAV-assisted network which aims to maximize the number of covered UEs while minimizing the interference between UAVs. \textcolor{black}{Moreover, the movement control for the UAVs serving wireless communications has been studied by reference~\cite{Kim2018}.}

Nowadays, machine learning techniques have gained popularity in solving UAV deployment \textcolor{black}{problem} and the reinforcement learning algorithm \textcolor{black}{has proved} to be an efficient solution of solving dynamic problem such as UAVs movement management in the UAV-assisted network \cite{Zhang2018,Bayerlein2018,Liu2019}. 
In particular, a machine learning framework based on Gaussian mixture model (GMM) and a weighted expectation maximization (WEM) algorithm to predict the locations of UAVs with the total power consumption minimized was proposed in~\cite{Zhang2018}. Furthermore, authors in~\cite{Bayerlein2018} studied a Q-learning based algorithm to find the optimal trajectory to maximize the sum rates of fixed location ground users for a single UAV base station (UAV-BS), assuming a random initial location. \textcolor{black}{Reference~\cite{Liu2019} proposed a Q-learning based movement design for multiple UAV-BSs. In addition, authors in~\cite{Chen2020} also demonstrated a reinforcement learning leveraged handover mechanism for cellular-connected drone system.}

\subsection{Our Contribution}
Despite \textcolor{black}{aforementioned works}, there \textcolor{black}{has been} few study on real-time movement strategy for UAVs to cope with the ground UEs' \textcolor{black}{mobility} in a UAV-assisted network, which is very critical in practical application scenarios. \textcolor{black}{In this research, we investigate a real-time dynamic UAV movement strategy design on a deep learning framework  called deep Q-network (DQN)~\cite{Volodymyr2015} to maximize the sum data rate. Unlike the existing literature \textcolor{black}{about UAV trajectory planning} in which the ground users are assumed geographically fixed, our contribution formulates the design problem of the UAVs' movement strategy to find the optimal locations of UAVs in every single time instant, in response to the ground users'} \textcolor{black}{random movement.}

\vspace{-0.25cm}
\section{System Model}
\label{sec:system_model}

\subsection{System Description}
\vspace{-0.0cm}

\begin{figure}[t!]
		\centering
		\includegraphics[width=0.75\columnwidth]{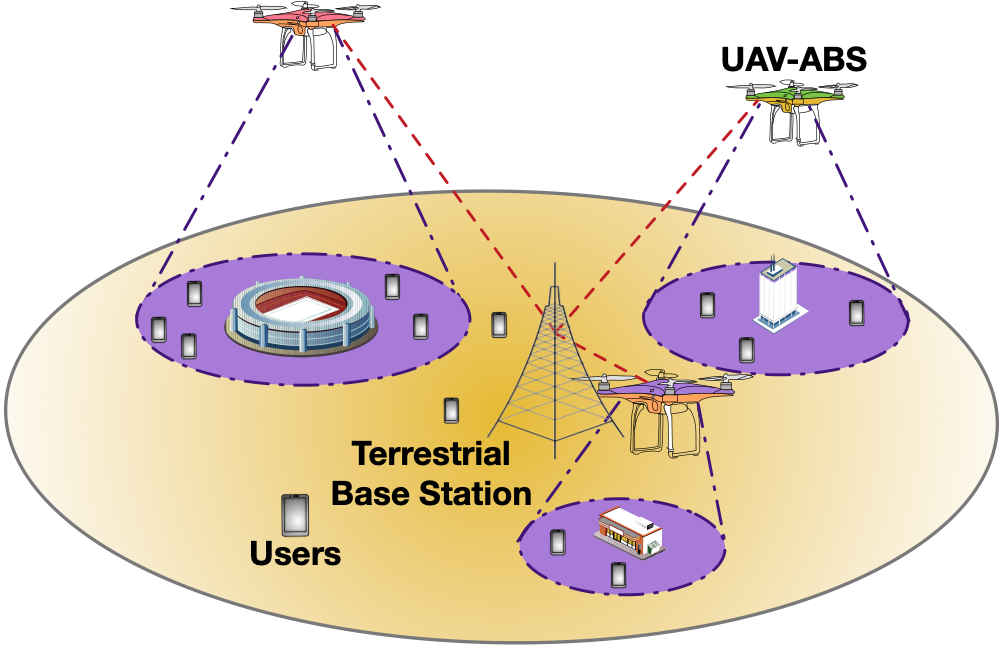}
		\caption{\textcolor{black}{A communication system model of UAV-assisted network.}}
		\label{fig:sys_mol_demo}
		\vspace{-0.4cm}
	\end{figure}

Fig. \ref{fig:sys_mol_demo} shows the framework of UAV-assisted wireless communications system model where UAVs serve as aerial base stations and provide hot spot wireless communications to the ground UEs. Also, the traditional terrestrial infrastructures are capable of serving the UEs which are not covered by UAV-BSs. Let $\mathcal P$ be the set of all the UEs which are labelled as $i = 1,2,...,\left | \mathcal P \right |$. $\mathcal Q$ denotes the set of available UAV-BSs labelled as $j = 1,2,...,\left | \mathcal Q \right |$ and $\mathcal O$ denotes the set of ground base stations (GBSs) labelled as  $k = 1,2,...,\left | \mathcal O \right |$. In our system, we assume that the UEs are assigned to the closest base station to receive wireless communication service and all the UAV-BSs cells are deployed at the same altitude $H$. 
Ground users are assumed to move from time to time and the location of the $i$-th UE at time $t$ can be expressed as $m_{i}(t)=\left [ x_{i}(t), y_{i}(t) \right ], t\in T$ where $T$ is the time window considered. Similarly, the locations of UAV-BS $j$ can be written as  $n_{j}(t)=\left [ \tilde x_{j}(t), \tilde y_{j}(t) \right ]$. Also, $u_{k}=\left [ \check x_{k}, \check y_{k} \right ]$ denotes the location of the $k$-th GBS, which is a known parameter in the study.

\subsection{Signal Model}
\vspace{-0.1cm}

The A2G channel model proposed in~\cite{Al-Hourani2014} considers the line-of-sight (LoS) \textcolor{black}{communication} occurring with a certain probability. At each time instant, the probability of \textcolor{black}{having} LoS and non line-of-sight (NLoS) \textcolor{black}{communication} between \textcolor{black}{the} UAV $j$ and \textcolor{black}{the} user $i$ are formulated in~\cite{Al-Hourani2014}
\begin{equation}
\begin{aligned}
\label{eqn:prob}
P_{LoS}=&\frac{1}{1+a\exp (-b(\frac{180}{\pi }\tan^{-1}(\frac{H}{r_{ij}})-a))},\\
P_{NLoS}=&1-P_{LoS},
\end{aligned}
\end{equation}
where $a$ and $b$ are environment dependent variables and $r_{ij} = \sqrt{(x_{j}-\tilde{x}_{i})^{2}+( y_{j}-\tilde{y}_{i} )^{2}}$ is the horizontal Euclidean distance between the $i^{th}$ user and $j^{th}$ UAV. Then the path loss (PL) for LoS and NLoS can be written as 
\begin{equation}
\begin{aligned}
\label{eqn:loss}
L_{LoS}=20\log (\frac{4\pi f_{c}d_{ij}}{c})+\eta _{LoS},\\L_{NLoS}=20\log (\frac{4\pi f_{c}d_{ij}}{c})+\eta _{NLoS},
\end{aligned}
\end{equation}
where $f_{c}$ is the carrier frequency, $c$ is the speed of light and $d_{ij}$ denotes the distance between the UE and UAV-BS given by $d_{ij}=\sqrt{H_{j}^{2}+r_{ij}^{2}}$. Moreover, $\eta _{LoS}$ and $\eta _{NLoS}$ are the environment dependent average additional PL for LoS and NLoS conditions, respectively. According to~\eqref{eqn:prob}-~\eqref{eqn:loss}, the PL can be written as
\begin{equation}
\begin{aligned}
\label{eqn:PL}
L_{ij}=&L_{LoS}\times P_{LoS}+L_{NLoS}\times P_{NLoS}\\
=&\frac{A}{1+a\exp (-b(\arctan \frac{H}{r_{ij}}-a))}\\&+20\log (H^{2}+r_{ij}^{2})
+B,
\end{aligned}
\end{equation}
where $A=\eta _{LoS}-\eta _{NLoS}$ and $B = 20\log (\frac{4\pi f_{c}}{c})+\eta _{NLoS}$.

The path loss for UEs which are associated with the GBSs at distance $r_{ik}$ can be modeled by $L_{ik}=\eta r_{ik}^{\alpha}$ where $\eta$ is the additional PL over the free space PL and $\alpha$ is the PL exponent.

Moreover, the signal-to-interference-plus-noise ratio (SINR) experienced at a UE at a distance $r_{ij}$ from its associated UAV-BS $j$ can be expressed as

\begin{equation}
\begin{aligned}
\label{eqn:SINR}
SINR_{ij}=\frac{P_{j}h_{ij}L_{ij}^{-1}}{\sigma ^{2}+\sum_{\bar{j}\in \mathcal Q\setminus j}I_{i\bar{j}} + \sum_{k\in \mathcal O}I_{ik}},
\end{aligned}
\end{equation}
where
\begin{equation}
\begin{aligned}
\label{eqn:I}
I_{i\bar{j}}=P_{\bar{j}} h_{i\bar{j}}PL^{-1}_{i\bar{j}}~\text{and}~
I_{ik}=P_k h_{ik} PL_{ik}^{-1},
\end{aligned}
\end{equation}
represents interference from other UAV-BSs/GBSs, $P_{j}$ represents the transmit power of the $j^{th}$ base station, $h_{ij}$ is the small fading power assumed to be an independent number following the exponential distribution and $\sigma ^{2}$ is the variance of the additive white Gaussian noise component. For the UEs served by a GBS, their SINRs can be expressed in a similar manner. According to the Shannon Capacity Theorem, the data rate $C_{i}$ of the $i^{th}$ UE can be expressed as $C_{i}=B\log_{2}(1+SINR_{ij})$ where $B$ is the bandwidth of the channel.

\vspace{-0.2cm}
\section{Fundamental of Reinforcement Learning}
\label{sec:RLproblem}
Reinforcement learning \textcolor{black}{generally} proceeds in a cycle of
interactions between an agent and its environment. At time $t$, the agent observes a state $s_{t}\in S$, \textcolor{black}{and} performs an action $a_{t}\in A$ and subsequently receives a reward $r_{t}\in R$. The time index is then incremented and the environment propagates the agent to a new state $S_{t+1}$, from where the cycle restarts. \textcolor{black}{Therefore, the whole process is a Markov Decision Process (MDP)~\cite{Altman1999}.}

The task of the reinforcement learning is to train an agent \textcolor{black}{interacting} with the environment to provide the feedback to each of its actions. The agent arrives at different states by performing actions \textcolor{black}{that} lead to a reward so \textcolor{black}{that we could} reinforce the agents to learn to choose the best actions based on the reward. Therefore, the only objective of the agent is to maximize its total reward across an episode. The way the agent chooses its actions is known as policy.

\subsection{Q-Learning}
Q-learning specifically allows an agent to learn to act optimally in a given environment. The goal for the agent is to learn a behavior rule that
maximizes the reward it receives. 
Q-learning is an off-policy reinforcement learning algorithm which finds the best action for a given state. It is considered off-policy because the Q-learning function learns from actions that are outside the current policy. More specifically, Q-learning learns a policy that maximizes the total reward.

\begin{itemize}
\item Q-Value: The Q-Value $Q(s,a)$ represents the total rewards of agents being at state $s$ and performing action $a$, the Q-Value for each state and action can be found in the Q-Table. It can be computed by:
  \begin{equation}
\begin{aligned}
\label{eqn:PL}
Q(s,a)=r(s,a)+\gamma max_{a}Q({s}',a)
\end{aligned}
\end{equation}
where the above equation states that the Q-Value which is derived from the agent being at state $s$ and taking action $a$ equals to the immediate reward $r(s,a)$ plus the highest possible Q-Value of the next state ${s}'$ times a discount factor $\gamma$ which represents the contribution of future rewards. \textcolor{black}{To be more specific, the Q-Value is the sum of the instantaneous reward at the current time step and an observation of the next time step.}

  \item Q-Table: Q-Table is a look up table which states the Q-Value that represents the future values of actions for each states, and is updated regularly.
  
\end{itemize}

To begin with, the Q-Table is initialized with all zeros. Then the agent chooses an action based on epsilon greedy strategy $\alpha$ that 90\% the agent chooses the action with highest Q-Value while 10\% the agent chooses a random action. Based on the action the agent chooses, the reward of performing the action is observed. \textcolor{black} {Then the updated Q-Value is calculated by the old Q-Value plus the learning rate times the sum of the instantaneous reward plus the expected future value. The expected future value equals the difference between the old Q-Value and maximum possible Q-Value for the next time step. The formula is shown as: }

  \begin{equation}
\begin{aligned}
\label{eqn:bellman}
Q_{new}(s,a)&=Q_{old}(s,a)+ \\
&\alpha(r(s,a)+\gamma maxQ({s}',a)-Q_{old}(s,a)).
\end{aligned}
\end{equation}

\subsection{Deep Q-Network}
The Q-Learning is a powerful algorithm to create a look up table for the agent so that the agent is capable for making rational action in each state. However, the drawback of Q-Learning is when there are too many states in the environment, it requires a large amount of memory since we need a long Q-Table. Therefore, the neural network is a powerful tool that can be utilized to compute Q-value.
	
In deep Q-Network, the next action is determined by the maximum output of the neural network. Referring to equation (\ref{eqn:bellman}), if we make the loss function $Loss=(r +\gamma max_{a}\tilde{Q}({s}',a;\Theta)-Q(s,a;\Theta ))^{2}$ where $\Theta$ represents the parameters of the Q-Network, it becomes a simple regression problem. 

However, in this loss function, $Q(s,a;\Theta )$ plays the role of a desired target in a regression problem which needs to be stationary in order to converge the network. Therefore a separate network is used to calculate the target. This target network has the same architecture as the network to predict Q-Value but with frozen parameters. The parameters of the predicted network are copied to target network in every $C$ iterations and $C$ is a predetermined value. 

Also, another important factor in Deep Q-Network is experience replay. It stores a fixed size of samples from training data into a memory tuple. In each training step, a mini-batch of samples are randomly selected from the memory to train the Q-Network. Experience replay breaks up the correlation in the training data by sampling batch of experiences randomly from a large memory pool which also helps the network to converge.

\vspace{-0.2cm}
\section{UAV-BS movement strategy}
\label{sec:problem}

The dynamic UAV-BS movement strategy problem can be treated as a design of determining the positions of the UAV-BSs at each time instant. The objective is to find the optimal positions for all UAV-BSs in each time-slot, to maximize the sum data rates of users. $\gamma _{ij/ik}(t)$ is a binary variable indicating whether the user $i$ is associated with UAV-BS $j$ or GBS $k$ at time instant $t$, with 1 for service and 0 for no association. Thus, the optimization problem at each time instant $t$ can be formulated as:


\begin{equation}
	\begin{aligned}
	\label{eqn:optimization}
	&\underset{n_{j}(t),j\in \mathcal Q}{\text{maximize }}\sum_{i=1}^{\left | \mathcal P \right |}C_{i}(t),\\
	\text{s.t. }
	C1:&\left \| n_{j}(t)-\gamma _{ij}(t)m_{i}(t) \right \|\leq \left \| n_{\bar{j}}(t)-m_{i}(t)\right \| \\ &+M\left |1-\gamma _{ij}(t)  \right |,
	\forall j\in Q, \forall \bar{j}\in \left \{ \mathcal O,\mathcal Q\setminus j \right \}\\
	C2:&\left \|u_{k}-\gamma _{ik}(t)m_{i}(t) \right \|\leq \left \| u_{\bar{k}}-m_{i}(t)\right \| \\ &+M\left |1-\gamma _{ik}(t)  \right |,
	\forall k\in \mathcal O, \forall \bar{k}\in \left \{ \mathcal Q, \mathcal O\setminus k \right \}\\
	C3:&\sum_{j}\gamma _{ij}(t)+\sum_{k}\gamma _{ik}(t)=1,\forall i,j,k.
	\end{aligned}
\end{equation}

Constraints $C1$ and $C2$ in (6) guarantee all the UEs are associated with the nearest UAV-BSs/GBSs where, $M$ is a large number to ensure the constraints hold in any UE association conditions. Then $C3$ guarantees all the UEs are associated with only a single base station. Therefore, the objective of the optimization problem is to find the optimal positions of UAV-BSs in each instant over time duration $T$ so that the sum data rates of the users can be maximized. \textcolor{black}{Although the UEs' movement is random, the UEs' distribution follows a certain principle. As a consequence, the environment is considered as partly random and partly stationary, which makes the whole process follow the MDP.}

\vspace{-0.2cm}
\section{Deep Q-network based UAV-BSs movement design}
\label{sec:algorithm}
\begin{algorithm}[t!]
{\footnotesize\caption{\footnotesize Deep Q-Network Based UAV-BS Movement Strategy}
\label{alg:proposed_algorithm}
    \hspace*{\algorithmicindent} \textbf{Required: }Initial Position of UAV-BSs, $m_{i}(0)$ and UEs $n_{j}(0)$ \\
	\begin{algorithmic}[1]
		\STATE Initialize replay memory $D$ with capacity $N$, mini-batch size $B$, initialize action-value network $\bar{Q}$ with weight $\bar{\Theta }_{j\in Q}$, target network $\tilde{Q}$ with weight $\tilde{\Theta }_{j\in Q}$ with random weights. 
		\FOR {each episode}
		\STATE Reset UAV-BSs to the initial positions
		\FOR {each time step $t$}
		\FOR {each UAV-BS agent $j$}
		\STATE Observe $s_{t}^{(j)}$
		\STATE Choose the action $a_{t}^{j}$ which maximizes the $\bar{Q}(s_{t}^{j},a_{t}^{j};\bar{\Theta }_{j})$
		\ENDFOR
		\STATE All agents take actions, observe rewards $r_{t}^{j}$, update state $s_{t}^{j}\rightarrow s_{t+1}^{j}$
		\FOR {each UAV-BS agent $j$}
		\STATE Observe $s_{t+1}^{j}$
		\STATE Store ($s_{t}^{j}$,$a_{t}^{j},r_{t}^{j},s_{t+1}^{j}$) into replay memory $D_{j}$
		\STATE Uniformly sample mini batch from replay memory $D_{j}$

        \STATE Perform a gradient descent on $Loss = (r_t^{j} +\gamma max_{{a}'}\tilde{Q}(s_{t+1}^{j},{a}';\tilde{\Theta }_{j})-\bar Q(s_{t}^{j},a_{t}^{j};\bar \Theta_{j} ))^{2}$ with respect to network parameters $\bar \Theta_{j}$.
        \STATE Update $\tilde{\Theta }_{j}=\bar \Theta_{j}$ every $C$ time steps
		\ENDFOR
		\ENDFOR
		\ENDFOR

\end{algorithmic}}
\end{algorithm}
In this section, given the real-time locations of a set of UEs, we present a reinforcement learning based UAV-BS movement strategy to obtain the optimal real-time locations of UAV-BSs. Before discussing the movement of UAV-BSs, the mobility model of UEs needs to be discussed first. The random walk model~\cite{Ren2017} is chosen as the UE mobility model in this paper, but other models can be easily included. The moving direction of UEs are uniformly distributed among left, right, forward, backward and staying still. Moreover, the initial positions of the ground users are assumed to be fixed. At each instant $t\in T$ when ground users move, all UAV-BSs take action in response to the movement of the ground users. 

\begin{figure}[t!]
		\centering
		\includegraphics[width=0.95\columnwidth]{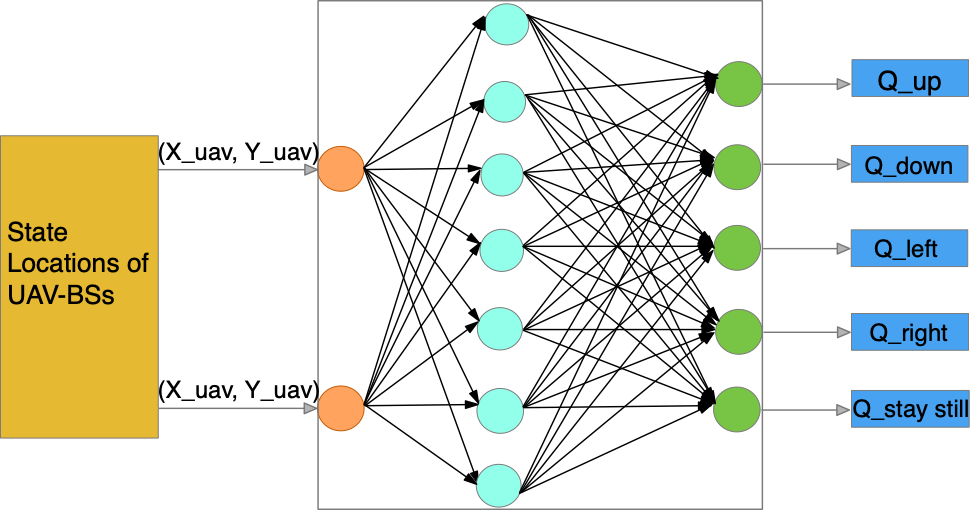}
		\caption{Deep Q-Network Structure.}
		\label{fig:struct}
		\vspace{-0.4cm}
	\end{figure}

The objective is to train a neural network to represent the action-value function which takes the local observations of the positions of both UEs and UAV-BSs in any instant as inputs and derives the action-value functions of the UAV-BSs movement. The Deep Q-Network consists of four parts: states, actions, rewards and the Q-Network training which is illustrated in Fig. \ref{fig:struct}. At each time slot $t$, each agent observes a state $s_{t}$, from the state space $S$ and takes an action $a_{t}$ in the action space $A$ based on the decision from Q-Network $\bar Q$. The principle of the Q-Network is to obtain the maximum Q-value which maximizes the sum data rates of UEs. Following the action, the state of each agent transits to a new state $s_{t+1}$ and the agents receive a reward $r_{t}$ which is determined by the instantaneous sum data rates of ground users. 

\subsection{State Representation}
\vspace{-0.0cm}

All agents' states are defined as: $s =(x_{uav},y_{uav})$ which is the horizontal position of the UAVs. Assuming that the initial states of all UAV-BSs are at the optimal positions where the sum data rates of ground users are maximized at time instant $t_{0}$. The optimal positions can be derived by conducting exhaustive search.

\subsection{Environment}
\vspace{-0.0cm}
\textcolor{black}{The Deep Q-Network addresses constraints by responding to the actions taken by the agents from the environment. In our problem formulation, the UEs are assigned to the nearest base stations and one UE can only be assigned to one base station. Therefore, the feedback from the environment has to follow these constraints to decide the instantaneous rewards. 
}

\subsection{Action Space}
\vspace{-0.0cm}

At each time step, all the UAV-BSs take an action $a_{t}\in A$ which includes choosing a direction for UAV-BSs to move according to the current state $s_{t}$, based on the decision from Q-Network $\bar Q$. In our model, we assume that all UAV-BSs move in the same speed in any time step, therefore the moving distance for any UAV-BS from any time instant $t$ to $t+1$ is assumed to be the same. More specifically, since we assume that all the UAV-BSs are at the same altitude $H$, there are 5 different actions in $A$: (1,0) means the UAV-BS will turn right, (-1,0) means the UAV-BS will turn left, (0,1) means the UAV-BS will move forward, (0,-1) means the UAV-BS will move  backward and (0,0) means the UAV-BS will stay still. \textcolor{black}{All UAV-BSs take actions one after another in a sequential manner.}  

\subsection{Reward Design}
After performing an action, the UAV-BS has a different location so the UEs need to change the association based on problem (\ref{eqn:optimization}). Therefore, the new association comes with a new instantaneous sum data rates of the ground UEs. The principle of designing the reward function is to improve the UEs' instantaneous data rates, which enables the agent to receive a positive reward. When the action results in a reduction of the sum data rates of the UEs, the UAV-BS receives a negative reward. Thus, the reward function can be expressed as
		\begin{equation}
        r_{t}=\left\{
        \begin{array}{rcl}
        1, & & {\text{if sum rates increase}}\\
        -0.2, & & {\text{if sum rates remain the same,}}\\
        -1,& & {\text{if sum rates decrease}}
        \end{array} \right.
        \end{equation}
\vspace{-0.0cm}
\textcolor{black}{where the ratio of the positive reward and negative reward is 1 in order to avoid any bias. Also, the reason to design a reward for an unchanged sum data rate is that in a practical situation, moving an UAV consumes the energy and resource, if the movement does not contribute to the objective, a negative reward is granted but its absolute value is much less than the case of a decreased sum data rate.}

\subsection{Training Procedure}
The training procedure requires a learning rate $\alpha $ and a discount factor $\gamma $. The learning procedure is divided into several episodes, and the positions of UAV-BSs will be reset to the initial values at the beginning of each episode. We leverage a DQN with experience replay to train the agents~\cite{Volodymyr2015}. In each episode, each agent takes actions based on the Q-Values which are outputted by the neural network and a reward is generated in each step. Therefore, the parameters of the neural network can be updated so after going through all the training episodes the neural network is capable of rational decisions for the UAV-BSs for each step. To be more specific, each agent $j$ has a DQN $\bar{Q}$ that takes an input of the observation of the current state $s_{t}^{j}$ and generate the output of the value functions corresponding to all the actions. At each training step $t$, each agent chooses the action $a_{t}^{j}$ which leads to the maximum estimated Q value. Based on the action taken by the agent, the transition tuple $(s_{t}^{j}, a_{t}^{j},r_{t}^{j},s_{t+1}^{j})$ is collected and stored into the replay memory $D$ with a size of $N$. Then, in each episode, a predetermined size of the mini-batch experiences $E$ are uniformly sampled to update $\Theta $ using gradient descent method to minimize the loss function
\begin{equation}
\begin{aligned}
\label{eqn:LOSS}
Loss=\sum_{E}(r_{t}^{j} +\gamma max_{{a}'}\tilde{Q}(s_{t+1}^{j},{a}';\tilde{\Theta }_{j})-\bar Q(s_{t}^{j},a_{t}^{j};\bar \Theta_{j} ))^{2}
\end{aligned}
\end{equation}
where $\tilde{\Theta }_{j}$ is the parameter set of a target network $\tilde Q$ which is replaced by the parameter set $\bar{\Theta }_{j}$ of training Q-Network $\bar Q$ every $C$ time steps. The experience replay can improve the training efficiency by breaking the correlation between samples so as to stabilize the training. 

\vspace{-0.2cm}
\section{numerical results}
\label{sec:experiment}

\begin{figure}[t!]
		\centering
		\includegraphics[width=0.95\columnwidth]{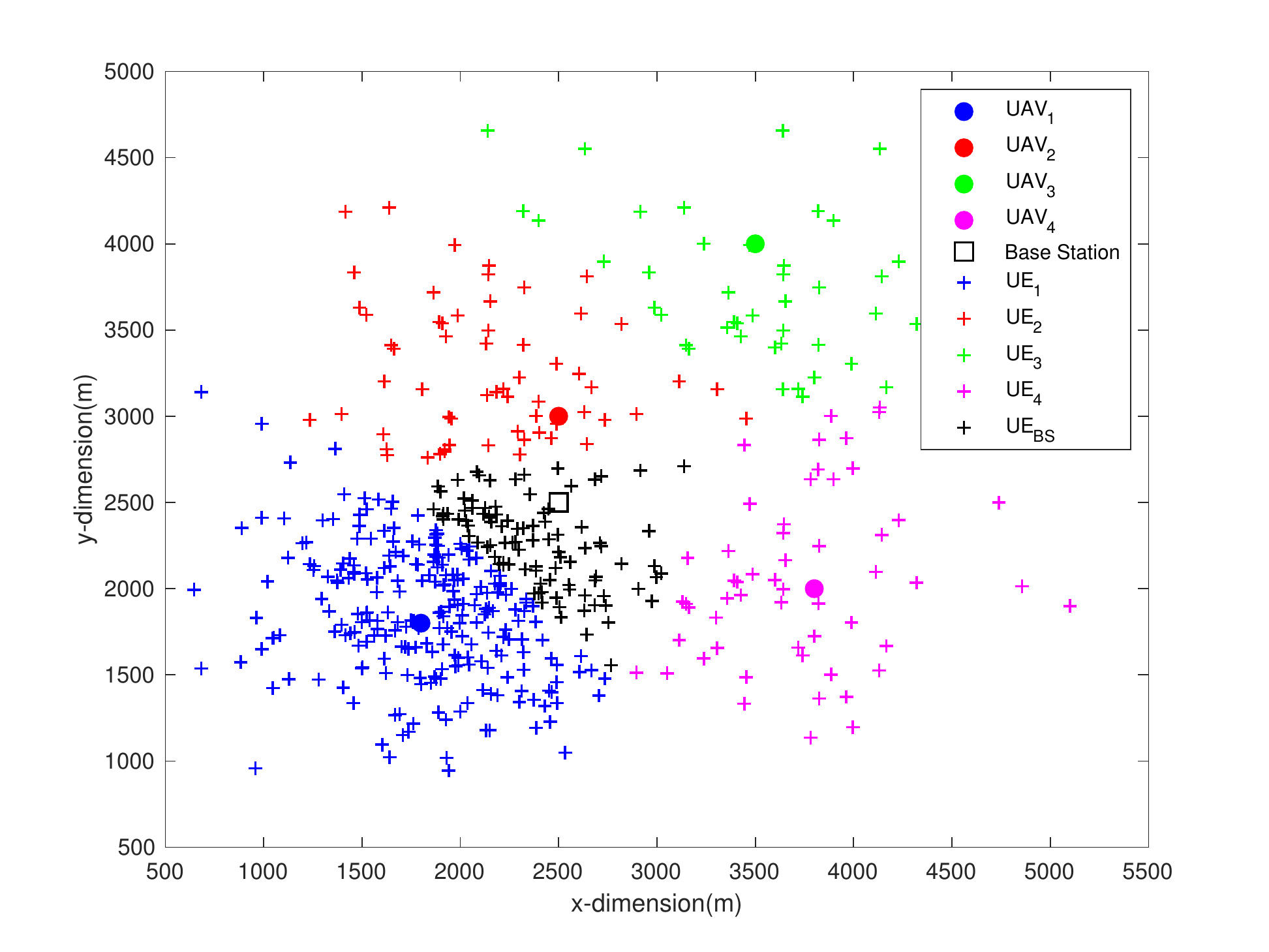}
		\caption{Snapshot of 500 UEs and association with 4 UAV-BSs in a 5000 m $\times $ 5000 m area.}
		\label{fig:location}
		\vspace{-0.4cm}
	\end{figure}

In our simulation, we consider UAV-assisted model in a 5000 m $\times $ 5000 m area and uniformly divide the entire area into 4 sections, \textcolor{black}{i.e.}, Section 1 : ${0<x\leq 2500, 0<y\leq 2500}$, Section 2 : ${2500<x\leq 5000, 0<y\leq 2500}$, Section 3 : ${0<x\leq 2500, 2500<y\leq 5000}$, Section 4 : ${2500<x\leq 5000, 2500<y\leq 5000}$. We assume that initially all of the UEs are distributed in the whole area, and then in the middle of the time duration, the majority (90\%) of the UEs converge to Section 1. At the end of the time duration, all the UEs go back to the uniformly distributed in the whole area. The UEs follow random walk mobility model inside the section area. There is one GBS available located at $u_{0}=[2500,2500]$. Further, we consider a period $T$ with 500 time instant and 50000 training episodes. Moreover, referring to~\cite{Al-Hourani2014}, the environment parameters are set up as follows: $f_{c}$ = 2 GHz, $PL_{max}$ = 103 dB, ($a$, $b$, $\eta _{LoS}$, $\eta _{NLoS}$) is configured to be (9.61, 0.43, 0.1, 20) corresponding to the urban environment. The transmit powers of UAV-BSs and GBS are set to be 37 dBm and 40 dBm, respectively. Also, the Deep Q-Network parameter set ($\alpha ,\beta ,N, B, C$) is configured to be (0.01, 0.9, 2000, 50, 200) and the structure of the network is configured to be 2 input neurons in the input layer, 10 neurons in the hidden layer and 5 neurons in the output layer. Also, the movement step size for UAV-BS is configured to be 1 meter. Fig. \ref{fig:location} shows the UEs distribution and their association in one time instant. The UEs and base stations with same color represent the association and all the UEs are associated with the closest base stations.
	
			\begin{figure}[t!]
		\centering
		\includegraphics[width=0.95\columnwidth]{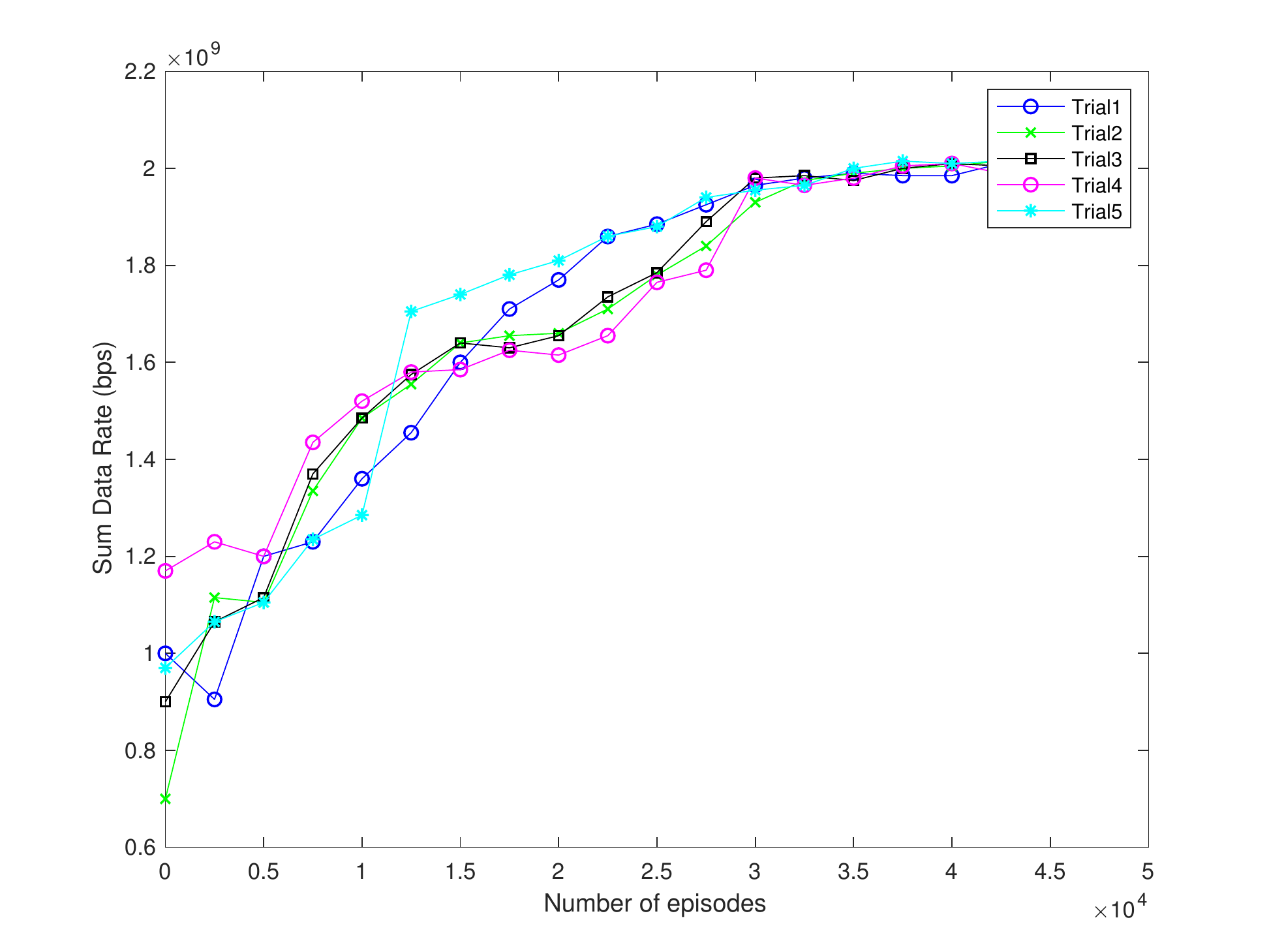}
		\caption{Sum data rate versus \textcolor{black}{the} number of training episodes.}
		\label{fig:train}
		\vspace{-0.4cm}
	\end{figure}
	
		\begin{figure}[t!]
		\centering
		\includegraphics[width=0.95\columnwidth]{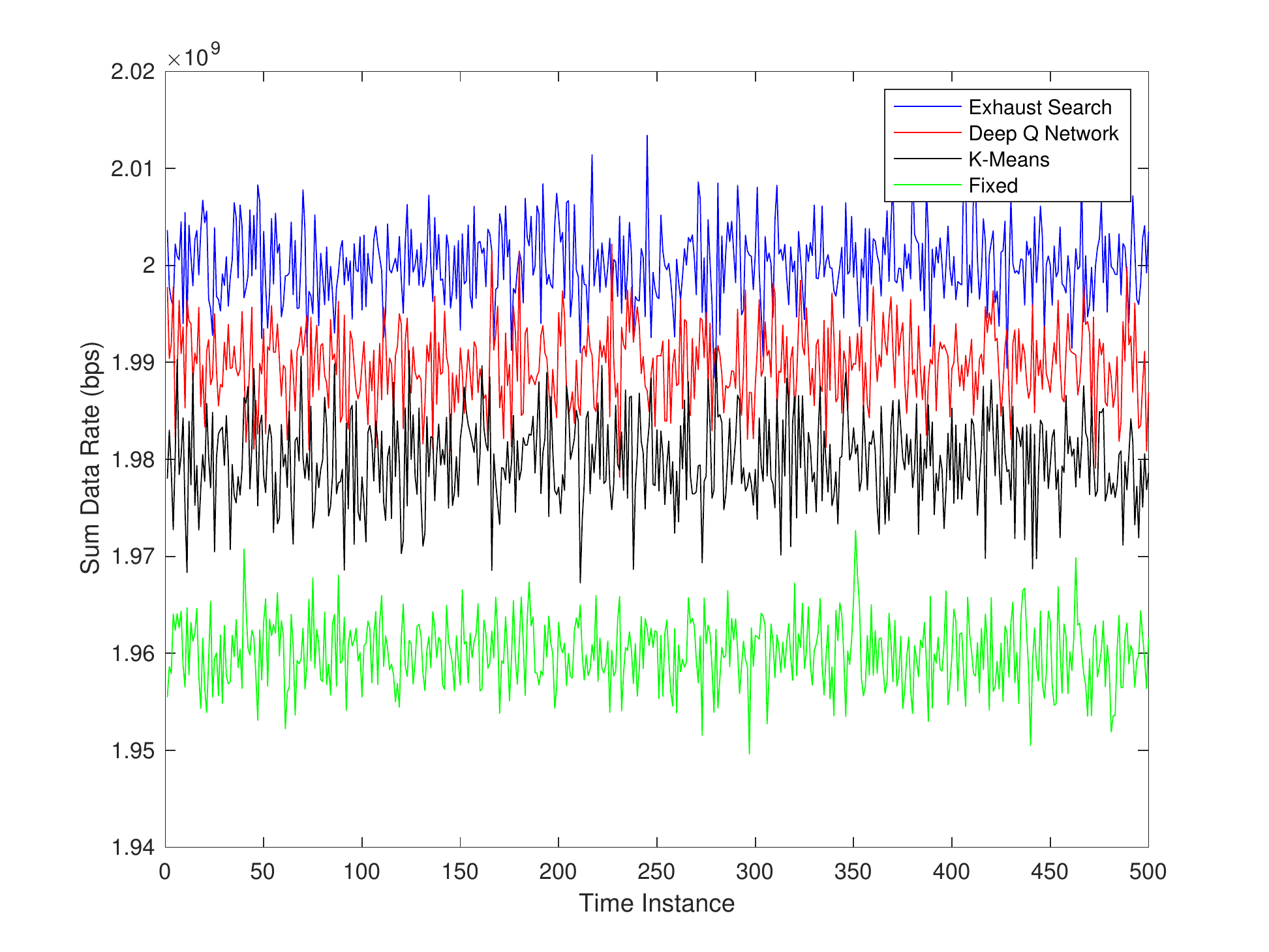}
		\caption{Sum data rate comparison of different methods.}
		\label{fig:final_plot}
		\vspace{-0.4cm}
	\end{figure}

	\begin{table}[h!]
\vspace{0cm}
\centering
\caption{Comparison of processing time of different algorithms}
\begin{tabular}{|c c|} 
 \hline
NA & Processing Time (ms)  \\ [-0.5ex] 
 \hline\hline
 Deep Q-Network &  210 \\ 
 Exhaustive Search & 4117  \\
 K-Means & 387  \\
 Fixed  &  0 \\
 [0ex] 
 \hline
\end{tabular}
\label{table:1}
\vspace{-0.4cm}
\end{table}

Fig. \ref{fig:train} further plots the sum data rates against the number of training episodes. It can be observed that the UAV-BSs are capable of carrying out their actions via iterative learning from their past experience to improve the performance.

Fig. \ref{fig:final_plot} shows the comparison of the sum data rates in all the time instants with different algorithms. It can be observed that the overall performance in 500 time instant of Deep Q-Network outperforms the fixed locations or K-Means deployment strategy and closely follows the performance of the exhaustive search. However, considering the computation cost results in Table \ref{table:1}, which is obtained using Intel® Core™ i5- 4430 Processor to run the algorithm 10 times and take the average processing time. Exhaustive search as expected achieves the highest performance but the computation complexity can be too high for real-time processing. The Deep Q-Network performs close to the exhaustive search but with significantly less processing resource and time, which is particularly critical for low-latency communications and mission execution involving UAVs.


\vspace{-0.25cm}
\section{Conclusion}
\label{sec:conclusion}
This paper has proposed and evaluated a dynamic UAV-BS deployment strategy for optimizing the real-time performance of wireless communication services when all the UEs are moving. A Deep Q-Network based algorithm has been proposed to maximize the sum data rates of ground UEs in a dynamic UAV-assisted network. Results have shown that the proposed algorithm outperforms other existing dynamic deployment algorithms.

\textcolor{black}{There is a potential direction for the future works. For example, in our research, we have considered a relatively large area where the time step is set to be relatively small, therefore, the action taken by one agent has little impact on the other UAV-BSs. Using a multi-agent reinforcement learning to enable multiple UAVs to take actions while considering their interactive impacts will be performed in the future study.}

\ifCLASSOPTIONcaptionsoff
  \newpage
\fi
\bibliographystyle{IEEEtran}
\bibliography{IEEEabrv,biblist}
\end{document}